\documentstyle[11pt]{article}

\def\cal{\ }
\def\Green{\ }

\def\Magenta{\ }

\def\Purple{\ }
\def\cal{\mathcal}

\begin{document}

\title{
\vspace*{-0.40in}
\begin{flushright}
{\normalsize UCTP-108-02}
\end{flushright}
\vspace*{0.10in}
Evidence for Light Scalar Resonances 
in Charm Meson Decays from Fermilab E791 }

\author{A.~J.~Schwartz \\
(\,for the E791 Collaboration\footnote{\uppercase{T}he \uppercase{E}791 
\uppercase{C}ollaboration is:
\uppercase{CBPF} (\uppercase{B}razil),
\uppercase{UC S}anta \uppercase{C}ruz,
\uppercase{C}incinnati,
\uppercase{CINVESTAS} (\uppercase{M}exico),
\uppercase{F}ermilab,
\uppercase{IIT} (\uppercase{C}hicago),
\uppercase{K}ansas \uppercase{S}tate, 
\uppercase{M}assachusetts,
\uppercase{M}ississippi,
\uppercase{P}rinceton,
\uppercase{P}uebla (\uppercase{M}exico),
\uppercase{S}outh \uppercase{C}arolina,
\uppercase{S}tanford, 
\uppercase{T}el \uppercase{A}viv,
\uppercase{T}ufts,
\uppercase{W}isconsin, and
\uppercase{Y}ale.}
\,) \\
{\it Department of Physics} \\
{\it University of Cincinnati} \\ 
{\it Cincinnati, OH, 45221}
}
\maketitle

\abstract{
From Dalitz-plot analyses of $D^+\rightarrow\pi^-\pi^+\pi^+$
and $D^+\rightarrow K^-\pi^+\pi^+$ decays, we find evidence for
light and broad scalar resonances $\sigma(500)$ and $\kappa(800)$.
From a Dalitz-plot analysis of $D^+_s\rightarrow\pi^-\pi^+\pi^+$ 
decays, we measure the masses and decay widths of the scalar
resonances $f^{}_0(980)$ and $f^{}_0(1370)$.
}

\section{Introduction}

The constituent quark model of QCD provides a successful
description of pseudo-scalar and vector mesons, which can
be conveniently arranged into $SU(3)^{}_c$ nonet 
representations. One also expects a 
corresponding scalar meson nonet to exist, and much experimental
effort has been directed towards identifying its members. The 
following states have been identified thus far, listed by isospin:
\begin{itemize}
\item $I\!=\!0$: $f^{}_0(600)$ or $\sigma(500)$, 
$f^{}_0(980)$, $f^{}_0(1370)$, $f^{}_0(1500)$, $f^{}_0(1710)$
\item $I\!=\!1/2$: $\kappa(800)$, $K^*_0(1430)$
\item $I\!=\!1$: $a^{}_0(980)$, $a^{}_0(1450)$
\end{itemize}
Scalar mesons have traditionally been studied via scattering experiments.
However, in these experiments the mesons can be difficult to disentangle 
from nonresonant background due to their broad widths and lack of a distinctive 
angular distribution. Thus some scalar states are controversial 
($\sigma(500)$ and $\kappa(800)$), while others, although accepted, 
have poorly-determined parameters ($f^{}_0(980)$, $a^{}_0(980)$, and 
$f^{}_0(1370)$). For a recent review, see Ref.~\cite{theory}.

Recently, large samples of $D^+_{(s)}\!\rightarrow\!\pi^-\pi^+\pi^+$
and $D^+\!\rightarrow\!K^-\pi^+\pi^+$ decays have been used to study 
scalar mesons ($S$) via the quasi-two-body decay
$D^+\!\rightarrow\!S\pi^+,\ S\!\rightarrow\!h^-\pi^+$
(charge-conjugate modes are assumed unless noted otherwise). 
The scalar masses and decay widths are determined by fitting 
a Dalitz-plot distribution to the square of coherently-summed decay 
amplitudes $|\sum_n {\cal A} (D\!\rightarrow\!S^{}_n\pi)|^2$. In this 
paper, we present such results from Fermilab E791. This experiment 
produced $D$ mesons using a 500~GeV/$c$ $\pi^-$ beam incident on Pt
and C targets. The subsequent $D$ decays were reconstructed using a 
silicon-strip vertex detector, a large-aperture spectrometer, 
and various particle identification detectors. 
%The experiment ran with a loose transverse energy trigger, 
%recording more than $2\times 10^9$ events. 
Details of the experiment can be found in Ref.~\cite{e791}.

\section{Dalitz-Plot Formalism}

The analyses select $D^+_{(s)}\!\rightarrow\!\pi^-\pi^+\pi^+$
and $D^+\!\rightarrow\!K^-\pi^+\pi^+$ decays by requiring that
there be a 3-track vertex well-separated from the primary 
interaction vertex. Pions and kaons are identified using
information from two threshold \u{C}erenkov counters. The 
$D^+\!\rightarrow\!K^-\pi^+\pi^+$ sample contains 15090 events 
with about 6\% background, while the $D^+\!\rightarrow\!\pi^-\pi^+\pi^+$ 
and $D^+_s\!\rightarrow\!\pi^-\pi^+\pi^+$ samples contain 1686 
and 937 events, respectively, with about 30\% background.
The $K\pi\pi$ sample is significantly larger and has a lower
level of background because this decay is Cabibbo-favored.

To study the resonance structure of each sample, an unbinned
maximum likelihood fit is performed. The likelihood function is 
${\cal L}=\prod^{}_{events} \left[ {\cal P}^{}_S + {\cal P}^{}_B\right]$,
where ${\cal P}^{}_S$ is the probability density function (PDF) for 
signal and ${\cal P}^{}_B$ is the PDF for background. The latter 
PDF is obtained from a fit to events in mass sidebands
or from Monte Carlo simulation. The signal PDF is 
${\cal P}^{}_S = (1/N^{}_S)\,g(M)\,\varepsilon(m^2_{12},m^2_{13})\,|{\cal A}|^2$,
where $N^{}_S$ is a normalization factor, $g(M)$ describes the signal
shape in the $\pi\pi\pi$ or $K\pi\pi$ mass spectrum, and 
$\varepsilon(m^2_{12},m^2_{13})$ is the acceptance over the
Dalitz plot, including smearing. The amplitude ${\cal A}$ is
the coherent sum of a uniform non-resonant (NR) amplitude and
amplitudes of resonances: 
\begin{equation}
{\cal A} = a^{}_0 e^{i\delta^{}_0}{\cal A}^{}_0 + \sum_{n=1}^{N}
a^{}_n e^{i\delta^{}_n}{\cal A}^{}_n(m^2_{12},m^2_{13})\,.
\end{equation}
The coefficients $a^{}_n$ and relative phases $\delta^{}_n$
are determined from the fit. The amplitude 
${\cal A}^{}_n = F^{(J)}_D\,F^{(J)}_n\,{\cal M}^{(J)}_n\,BW^{}_n$,
where $F^{(J)}_D$ and $F^{(J)}_n$ are Blatt-Weisskopf factors
that depend on the spin ($J$) and radii ($r^{}_D$ and $r^{}_R$) 
of the parent $D$ and intermediate resonance; 
${\cal M}^{(J)}_n$ is the angular-momentum-conserving factor
$(-2)^J p^J q^J P^{}_J(\cos\theta)$, where $\vec{p},\ \vec{q}$
are the momenta of the $D$ and one of the daughters of the
intermediate resonance in the resonance rest frame, and 
$\cos\theta = \hat{p}\cdot\hat{q}$;
and $BW^{}_n$ is the relativistic Breit-Wigner propagator
$\left[ m^2_n - m^2 - im^{}_n\Gamma^{}_n(m)\right]^{-1}$,
where $m$ is the invariant mass of the track pair forming
a resonance ($m^{}_{12}$ or $m^{}_{13}$), $m^{}_n$
is the resonance mass, and $\Gamma^{}_n(m)$ is the
mass-dependent width. More details of these expressions 
can be found in Ref.~\cite{d+_paper}.
Finally, each resonant amplitude is Bose-symmetrized
with respect to the identical pions:
${\cal A}^{}_n = {\cal A}^{}_n [({\bf 12}){\bf 3}] + 
{\cal A}^{}_n [({\bf 13}){\bf 2}]$.

\section{Results}

The results of the likelihood fits are given in 
Tables~\ref{tab:d+_results}--\ref{tab:kappa_results}. The 
tables list the coefficients $a^{}_n$ and phases $\delta^{}_n$ 
resulting from the fits, along with the decay fraction for each 
amplitude. The decay fraction for an amplitude is defined as 
its intensity integrated over the Dalitz plot, divided by the 
integrated intensity of all amplitudes coherently summed.
To assess the quality of the fit, a fast Monte Carlo 
program was developed that produces binned Dalitz-plot
densities according to signal and background PDFs,
including detector smearing effects. The difference 
between this density distribution and that of the data 
--\,summed over all bins\,-- is taken as the $\chi^2$
of the fit. The number of bins is the number of
degrees of freedom.

Table~\ref{tab:d+_results} gives the results for
$D^+\!\rightarrow\!\pi^-\pi^+\pi^+$. The left-most
column lists results obtained when fitting to established
resonances, while the right-most column lists results obtained
when an additional scalar resonance is included. For the
latter fit, the $\chi^2$ per degree of freedom ($\nu$)
is significantly improved, and the non-resonant fraction 
constitutes a much smaller component of the total width 
(in the left-most column it dominates). The mass and 
width obtained by the fit for the additional resonance
(denoted ``$\sigma$'') are 
$m=478^{+24}_{-23}\pm 17$~MeV/$c^2$
and $\Gamma=324^{+42}_{-40}\pm 21$~MeV/$c^2$.
As a test we also fit the data with vector,
tensor, and toy (Breit-Wigner with no phase variation) models for
the additional amplitude. In all cases we
obtain a higher $\chi^2/\nu$ than that obtained
with the additional scalar amplitude.

Table~\ref{tab:ds_results} gives the results for
$D^+_s\!\rightarrow\!\pi^-\pi^+\pi^+$. The left-most 
column lists results obtained when fitting to established
resonances, while the other two columns list results 
obtained when the $\rho^0(770)$ and $\rho^0(1450)$ 
are excluded. The parameters determined from the fit 
are the mass and width of the $f^0(1370)$, and the mass
and coefficients $g^{}_\pi,\ g^{}_K$ of
a coupled-channel Breit-Wigner width\,\cite{ds_paper} 
for the $f^0(980)$. The results are
$m^{}_{f^{}_0(1370)}=1434\pm 18\pm 9$~MeV/$c^2$,
$\Gamma^{}_{f^{}_0(1370)}=172\pm 32\pm 6$~MeV/$c^2$,
and $m^{}_{f^{}_0(980)}=977\pm 3\pm 2$~MeV/$c^2$, 
$g^{}_\pi=0.09\pm 0.01\pm 0.01$, 
$g^{}_K=0.02\pm 0.04\pm 0.03$. 
We also fit the data using the same Breit-Wigner 
for the $f^{}_0(980)$ as that used for the other resonances; 
the resulting fit is nearly as good and the decay fractions 
and phases are essentially unchanged.
The isoscalar plus $\pi^+$ components correspond to over
90\% of the $D^+_s\!\rightarrow\!\pi^-\pi^+\pi^+$ decay
width.

Table~\ref{tab:kappa_results} gives the results 
for $D^+\!\rightarrow\!K^-\pi^+\pi^+$. The left-most 
column lists results when fitting only to established
resonances and fixing the parameters of the scalar 
$K^*_0(1430)$. The middle column lists results obtained 
when the $K^*_0(1430)$ parameters are allowed to float, 
and, in addition, Gaussian form factors 
$F^{}_D\!\cdot\!F^{}_R = \exp(-p^2r^2_D/12)\!\cdot\!\exp(-p^2r^2_R/12)$
are used for the $K^*_0(1430)$ to account for the size of 
the decaying mesons. The meson radii $r^{}_D$ and $r^{}_R$ 
(also appearing in Blatt-Weisskopf factors) are now determined
from the fit. The right-most column lists results when an 
additional scalar resonance is included (also with 
Gaussian form factors in its amplitude). This last 
fit yields a significantly improved $\chi^2/\nu$, and 
the NR fraction is reduced from about 90\% to 13\%,
which is more consistent with expectations.
The mass and width of the additional 
resonance (denoted ``$\kappa$'') are $m=797\pm 19\pm 43$~MeV/$c^2$ 
and $\Gamma=410\pm 43\pm 87$~MeV/$c^2$. Including this additional
resonance, the fit obtains $K^*_0(1430)$ parameters  
$m^{}_{K^*(1430)}=1459\pm 7\pm 5$~MeV/$c^2$ and 
$\Gamma^{}_{K^*(1430)}=175\pm 12\pm 12$~MeV/$c^2$; these
values are shifted by $+47$~MeV and $-119$~MeV, respectively,
with respect to PDG values\,\cite{pdg}. The meson radii are 
$r^{}_D\!=\!5.0\pm0.5$~GeV$^{-1}$ and 
$r^{}_R\!=\!1.6\pm1.3$~GeV$^{-1}$.
We also fit the data with vector, tensor, and 
toy (Breit-Wigner with no phase variation) 
models for the additional amplitude; in all cases 
we obtain a higher $\chi^2/\nu$ than that obtained with 
the $\kappa$ amplitude.

In summary, we have performed Dalitz-plot analyses for the 
decays $D^+\!\rightarrow\!\pi^-\pi^+\pi^+$,
$D^+_s\!\rightarrow\!\pi^-\pi^+\pi^+$, and
$D^+\!\rightarrow\!K^-\pi^+\pi^+$.
For the first sample, including an additional scalar 
resonance with mass near 500~MeV/$c^2$ significantly improves the 
quality of the fit. For the $D^+\!\rightarrow\!K^-\pi^+\pi^+$ sample,
including a scalar resonance with $m\approx 800$~MeV/$c^2$ improves 
the quality of the fit. When this resonance is included, the fit 
obtains a mass and decay width for the $K^*_0(1430)$ that are 
shifted with respect to PDG values. 
For the $D^+_s\!\rightarrow\!\pi^-\pi^+\pi^+$ sample, the 
Dalitz-plot fit yields new values for the masses and decay 
widths of the $f^{}_0(980)$ and $f^{}_0(1370)$ scalar mesons.

\begin{table}
\begin{center}
\begin{tabular}{|c|cc|}
\hline
  & {\bf Estab.\ Resonances} &  {\bf {\boldmath Adding $\sigma(500)$}}  \\
\hline
  & Fraction (\%)  &  Fraction (\%)   \\
%  & Magnitude      &  Magnitude       \\
  & Phase          &  Phase           \\
\hline 
NR  & \Purple{$38.6\pm 9.7$} & \Magenta{$7.8\pm 6.0\pm 2.7$}     \\ 
%    & \Purple{$1.36\pm0.20$} & \Magenta{$0.48\pm 0.18\pm 0.09$}   \\
 & \Purple{$(150\pm 12)^\circ$} & \Magenta{$(57.3\pm 19.5\pm 5.7)^\circ$} \\
\hline
$\sigma\,\pi^+$	& \Purple{--}    & \Magenta{$46.3\pm 9.0\pm 2.1$}    \\
%		& \Purple{--}    & \Magenta{$1.17 \pm 0.13\pm 0.06$}  \\
		& \Purple{--}    & \Magenta{$(206\pm 8.0\pm 5.2)^\circ$} \\
\hline		     
$\rho^0(770)\pi^+$ 
  & \Purple{$20.8\pm 2.4$}  & \Magenta{$33.6\pm 3.2\pm 2.2$} \\
%  & \Purple{1 (fixed)} & \Magenta{$1$ (fixed)} \\
  & \Purple{$0^\circ$ (fixed)} & \Magenta{$0^\circ$ (fixed)} \\
\hline
$f_0(980)\pi^+$ 
  & \Purple{$7.4\pm 1.4$} & \Magenta{$6.2\pm 1.3\pm 0.4$} \\
%  & \Purple{$0.60\pm 0.07$} & \Magenta{$0.43\pm 0.05\pm 0.02$} \\
  & \Purple{$(152\pm 16)^\circ$} & \Magenta{$(165\pm 11\pm 3.4)^\circ$} \\
\hline
$f_2(1270)\pi^+$ 
  & \Purple{$6.3\pm 1.9$} & \Magenta{$19.4\pm 2.5\pm 0.4$}  \\
%  & \Purple{$0.55\pm 0.08$} & \Magenta{$0.76 \pm 0.06\pm 0.03$} \\
  & \Purple{$(103\pm 16)^\circ$} & \Magenta{$(57.3\pm 7.5\pm 2.9)^\circ$} \\
\hline
$f_0(1370)\pi^+$ 
  & \Purple{$10.7\pm 3.1$} & \Magenta{$2.3\pm 1.5\pm 0.8$} \\
%  & \Purple{$0.72\pm 0.12$} & \Magenta{$0.26\pm 0.09\pm 0.03$} \\
  & \Purple{$(143\pm 9.7)^\circ$} & \Magenta{$(105\pm 18\pm 0.6)^\circ$} \\
\hline		     
$\rho^0(1450)\pi^+$ 
  & \Purple{$22.6\pm 3.7$}  & \Magenta{$0.7\pm 0.7\pm 0.3$} \\
%  & \Purple{$1.04\pm 0.12$} & \Magenta{$0.14\pm 0.07\pm 0.02$} \\
  & \Purple{$(46\pm 15)^\circ$} & \Magenta{$(319\pm 39\pm 11)^\circ$} \\
\hline
$\chi^2/\nu$  & \Purple{254/162} & \Magenta{138/162} \\
\hline
\end{tabular}
\end{center}
\vskip0.10in
\caption{Results of the maximum likelihood fit to the
$D^+\!\rightarrow\!\pi^-\pi^+\pi^+$ Dalitz plot
(from Ref.~\cite{d+_paper}). \label{tab:d+_results} }
\end{table}

\begin{table}
\begin{center}
\begin{tabular}{|c|ccc|}
\hline
  & {\bf Estab.\ Resonances}  & {\bf {\boldmath No $\rho^0(770)$}}  &  
			\Magenta{\bf {\boldmath No $\rho^0(1450)$}}   \\
\hline
  & Fraction (\%)  & Fraction (\%)  &  \Magenta{Fraction (\%)}   \\
  & Magnitude      & Magnitude      &  \Magenta{Magnitude}       \\
  & Phase          & Phase          &  \Magenta{Phase}           \\
\hline 
NR  & $0.5\pm 1.4\pm 1.7$ & $7.5\pm 4.8$  & $5.0\pm 3.8$   \\ 
    & $0.09\pm 0.14\pm 0.04$ & $0.36\pm 0.12$ &  $0.30\pm 0.12$ \\
 & $(181\pm 94\pm 51)^\circ$ & $(165\pm 23)^\circ$ & $(149\pm 25)^\circ$ \\
\hline
$\bar f^{}_0(980)\pi^+$ 
  & $56.5\pm 4.3\pm 4.7$ & $58.0\pm 4.9$ & $54.1\pm 4.0$ \\
  & 1 (fixed)		  & 1 (fixed) & 1 (fixed) \\
  & $0^\circ$ (fixed) & $0^\circ$ (fixed) & $0^\circ$ (fixed) \\
\hline
$\rho^0(770)\pi^+$ 
  & $5.8\pm 2.3\pm 3.7$ & -- & $11.1\pm 2.5$ \\
  & $0.32\pm 0.07\pm 0.19$ & -- & $0.45\pm 0.06$ \\
  & $(109\pm 24\pm 5)^\circ$ & -- & $(81\pm 15)^\circ$ \\
\hline
$f^{}_2(1270)\pi^+$ 
  & $19.7\pm 3.3\pm 0.6$ & $22.2\pm 3.3$ & $20.8\pm 3.0$  \\
  & $0.59\pm 0.06\pm 0.02$ & $0.62\pm 0.06$ & $0.62\pm 0.05$ \\
  & $(133\pm 13\pm 28)^\circ$ & $(109\pm 11)^\circ$ & $(124\pm 11)^\circ$ \\
\hline
$f^{}_0(1370)\pi^+$  
  & $32.4\pm 7.7\pm 1.9$ & $30.4\pm 6.9$ & $34.7\pm 7.2$ \\
  & $0.76\pm 0.11\pm 0.03$  & $0.72\pm 0.11$ & $0.80\pm 0.11$ \\
  & $(198\pm 19\pm 27)^\circ$ & $(156\pm 19)^\circ$ & $(159\pm 14)^\circ$ \\
\hline
$\rho^0(1450)\pi^+$ 
  & $4.4\pm 2.1\pm 0.2$ & $5.8\pm 2.2$ & -- \\
  & $0.28\pm 0.07\pm 0.01$ & $0.32\pm 0.06$ & -- \\
  & $(162\pm 26\pm 17)^\circ$ & $(144\pm 20)^\circ$ & -- \\
\hline  
$\chi^2/\nu$  & 72/68 & 93/68 & 104/68 \\
\hline
\end{tabular}
\end{center}
\vskip0.10in
\caption{Results of the  maximum likelihood fit to the
$D^+_s\!\rightarrow\!\pi^-\pi^+\pi^+$ Dalitz plot
(from Ref.~\cite{ds_paper}).\label{tab:ds_results} }
\end{table}

\begin{table}
\hbox{\hskip-0.20in
\begin{tabular}{|c|ccc|}
\hline
 & {\bf Estab.\ Resonances} & {\bf {\boldmath Floating $K^*_0(1430)$ with}}
				 & {\bf {\boldmath Adding $\kappa(800)$}} \\
  &                & {\bf Gaussian form factors} &         \\
\hline
  & \Purple{Fraction (\%)}  & \Green{Fraction (\%)} &  \Magenta{Fraction (\%)}   \\
  & \Purple{Magnitude}      & \Green{Magnitude}     &  \Magenta{Magnitude}       \\
  & \Purple{Phase}          & \Green{Phase}         &  \Magenta{Phase}           \\
\hline 
NR  & \Purple{$90.9\pm 2.6$} & \Green{$89.5\pm 16.1$}  & \Magenta{$13.0\pm 5.8\pm 4.4$}   \\ 
    & \Purple{$1.0$ (fixed)} & \Green{$2.72\pm 0.55$} &  \Magenta{$1.03\pm 0.30\pm 0.16$} \\
 & \Purple{$0^\circ$(fixed)} & \Green{$(-49\pm 3)^\circ$} & \Magenta{$(-11\pm 14\pm 8)^\circ$} \\
\hline
$\kappa\,\pi^+$	& \Purple{--} & \Green{--} & \Magenta{$47.8\pm 12.1\pm 5.3$} \\
		& \Purple{--} & \Green{--} & \Magenta{$1.97 \pm 0.35\pm 0.11$} \\
		& \Purple{--} & \Green{--} & \Magenta{$(187\pm 8\pm 18)^\circ$} \\
\hline		     
$\bar K^*(892)\pi^+$ 
  & \Purple{$13.8\pm 0.5$} & \Green{$12.1\pm 3.3$} & \Magenta{$12.3\pm 1.0\pm 0.9$} \\
  & \Purple{$0.39\pm 0.01$} & \Green{1.0 (fixed)} & \Magenta{$1.0$ (fixed)} \\
  & \Purple{$(54\pm 2)^\circ$} & \Green{$0^\circ$ (fixed)} & \Magenta{$0^\circ$ (fixed)} \\
\hline
$\bar K^*_0(1430)\pi^+$ 
  & \Purple{$30.6\pm 1.6$} & \Green{$28.7\pm 10.2$} & \Magenta{$12.5\pm 1.4\pm 0.5$} \\
  & \Purple{$0.58\pm 0.01$} & \Green{$1.54\pm 0.75$} & \Magenta{$1.01 \pm 0.10\pm 0.08$} \\
  & \Purple{$(54\pm 2)^\circ$} & \Green{$(6\pm 12)^\circ$} & \Magenta{$(48\pm 7\pm 10)^\circ$} \\
\hline
$\bar K^*_2(1430)\pi^+$ 
  & \Purple{$0.4\pm 0.1$} & \Green{$0.5\pm 0.3$} & \Magenta{$0.5\pm 0.1\pm 0.2$}  \\
  & \Purple{$0.07\pm 0.01$} & \Green{$0.21\pm 0.18$} & \Magenta{$0.20 \pm 0.05\pm 0.04$} \\
  & \Purple{$(33\pm 8)^\circ$} & \Green{$(-3\pm 26)^\circ$} & \Magenta{$(-54\pm 8\pm 7)^\circ$} \\
\hline
$\bar K^*(1680)\pi^+$  
  & \Purple{$3.2\pm 0.3$} & \Green{$3.7\pm 1.9$} & \Magenta{$2.5\pm 0.7\pm 0.3$} \\
  & \Purple{$0.19\pm 0.01$}  & \Green{$0.56\pm 0.48$} & \Magenta{$0.45 \pm 0.16\pm 0.02$} \\
  & \Purple{$(66\pm 3)^\circ$} & \Green{$(36\pm 25)^\circ$} & \Magenta{$(28\pm 13\pm 15)^\circ$} \\
\hline
$\chi^2/\nu$  & \Purple{167/63} & \Green{126/63} & \Magenta{46/63} \\
\hline
\end{tabular}
}
\vskip0.10in
\caption{Results of the maximum likelihood fit to the
$D^+\!\rightarrow\!K^-\pi^+\pi^+$ Dalitz plot
(from Ref.~\cite{kappa_paper}).\label{tab:kappa_results} }
\end{table}


\begin{thebibliography}{0}

\bibitem{theory} F.\ E.\ Close and N.\ A.\ T\"{o}rnqvist,
{\it J.\,Phys.\/} {\bf G28}, R249 (2002); 
F.\ E.\ Close, {\it Int.\,J.\,Mod.\,Phys.\/} {\bf A17}, 3239 (2002).

\bibitem{e791} E.\ M.\ Aitala {\it et al.},
{\it Eur.\,Phys.\,J.\,direct\/} {\bf C1}, 4 (1999).

\bibitem{d+_paper} E.\ M.\ Aitala {\it et al.},
{\it Phys.\,Rev.\,Lett.\/} {\bf 86}, 770 (2001).

\bibitem{ds_paper} E.\ M.\ Aitala {\it et al.},
{\it Phys.\,Rev.\,Lett.\/} {\bf 86}, 765 (2001).

\bibitem{kappa_paper} E.\ M.\ Aitala {\it et al.},
{\it Phys.\,Rev.\,Lett.\/} {\bf 89}, 121801 (2002).

\bibitem{pdg} K.\ Hagiwara {\it et al.},
{\it Phys.\,Rev.\,D\/} {\bf 66}, 010001 (2002). 
See also note~[15] of Ref.~\cite{kappa_paper}. 

\end{thebibliography}
\end{document}